\documentclass[aps, prb, 10pt, superscriptaddress, twocolumn]{revtex4-1}

\usepackage{amsmath,amssymb}
\usepackage{graphicx}

\DeclareMathOperator{\re}{Re}

\allowdisplaybreaks[1]

\begin{document}

\title{Fisher--Hartwig expansion for the transverse correlation function\\ in the XX spin-1/2 chain}

\author{Dmitri A.~Ivanov}
\affiliation{Institute for Theoretical Physics, ETH Zurich, 8093 Zurich, Switzerland}
\affiliation{Institute for Theoretical Physics, University of Zurich, 8057 Zurich, Switzerland}

\author{Alexander G.~Abanov}
\affiliation{Department of Physics and Astronomy and Simons Center for Geometry and Physics,
Stony Brook University,  Stony Brook, NY 11794, USA}

\date{September 17, 2013}

\begin{abstract}
Motivated by the recent results on the asymptotic behavior of Toeplitz determinants
with Fisher--Hartwig singularities, we develop an asymptotic expansion for
transverse spin correlations in the XX spin-1/2 chain. The coefficients of
the expansion can be calculated to any given order using the relation to
discrete Painlev\'e equations. We present explicit results up to the eleventh order
and compare them with a numerical example.
\end{abstract}



\maketitle

\section{Introduction}
\label{sec:intro}

The spin-1/2 XX chain described by the Hamiltonian
\begin{equation}
H=\sum_{i=-\infty}^{+\infty} \left[-J(\sigma^x_i \sigma^x_{i+1} + \sigma^y_i \sigma^y_{i+1})
+ h\sigma^z_i \right]
\label{XX-hamiltonian}
\end{equation}
(where $\sigma^\alpha_i$ are Pauli matrices associated with the spin at the site $i$)
is one of the simplest examples
of exactly solvable spin systems: by a Jordan--Wigner
transformation it can be mapped onto free fermions,
which gives a complete solution for the ground state
and for the excitations \cite{lieb:61} (this model is also equivalent
to the Tonks--Girardeau gas of hard-core bosons on a one-dimensional lattice \cite{girardeau:60}). 
Computing correlation functions is, however, a more difficult task,
since transverse spin operators (or, equivalently, the boson
operators in the Tonks--Girardeau gas) are nonlocal in terms
of fermions. In particular, already the leading-order
asymptotic behavior of the ground-state transverse correlations
\begin{equation}
\langle \sigma^+_i \sigma^-_{i+L} \rangle \propto L^{-1/2}\, ,
\qquad L\to\infty
\label{CF-main-scaling}
\end{equation}
requires a nontrivial calculation \cite{BosLead}.  One of the possible approaches to
this correlation function is to re-express it as a Toeplitz
determinant of a Fisher--Hartwig type by using the fermionic 
representation \cite{lieb:61,schultz:63,mccoy:68}. The asymptotic behavior of such determinants
continues to be an object of active studies in mathematics and
mathematical physics \cite{fisher:68,basor:94,ehrhardt:01,basor:06,%
kozlowski:08:09:10,calabrese:10,gutman:11,deift:11,krasovsky:11}. 
In particular, some corrections
to the leading asymptotic behavior of the correlation function
(\ref{CF-main-scaling}) have been computed \cite{vaidya:79,jimbo:80,creamer:81,gangardt:04:06,note-vaidya}.

In the present paper we extend this approach by demonstrating that
all the corrections to Eq.~(\ref{CF-main-scaling}) can be computed
order by order using the recent results on a closely related 
Toeplitz determinant for statistics of free fermions \cite{ivanov:13-1,ivanov:13-2}. 
Furthermore, those corrections may be combined into a double sum explicitly periodic
in the ``counting parameter'' (Eqs.~(\ref{sigma-expansion-1})--(\ref{sigma-expansion-2}) below): 
this double-sum form was dubbed {\em Fisher--Hartwig expansion} in Ref.~\onlinecite{ivanov:13-2}.

Note that while our calculation can be performed analytically to
any order, it does not constitute a rigorous proof: in fact, the
results of Refs.~\onlinecite{ivanov:13-1,ivanov:13-2} on which it is based still have a status 
of conjecture. The available analytical
and numerical evidence \cite{calabrese:10,abanov:11,suesstrunk:12,ivanov:13-2} 
leaves no doubt in the validity of this
conjecture (see a more detailed discussion of this in Section 
\ref{sec:discussion}), however we find it helpful to check our analytical
results against numerical evaluations of the determinants.
Such a check provides an additional support to the conjecture
and verifies analytical manipulations performed with
Mathematica software \cite{Wolfram} (see Section \ref{sec:numerics} for detail).

The paper is organized as follows. In Section \ref{sec:results}, 
we report our analytical results 
on the Fisher--Hartwig expansion of the relevant Toeplitz determinant.
In Section \ref{sec:spin} we apply these results to the case of transverse
spin correlations in the XX chain.
In Section \ref{sec:numerics}, we compare the analytical results with a numerical example.
Finally, in Section \ref{sec:discussion} we discuss the assumptions used in our
calculation and the implications of our results. The Appendix contains
expressions for some of the expansion coefficients.

\section{Main results}
\label{sec:results}

The Hamiltonian (\ref{XX-hamiltonian}) can be diagonalized via the Jordan--Wigner 
transformation \cite{lieb:61}
\begin{align}
\sigma^+_i &= \Psi^\dagger_i \exp \left(i\pi \sum_{j<i} \Psi^\dagger_j \Psi_j \right)\, , \nonumber\\
\sigma^-_i &= \exp \left(i\pi \sum_{j<i} \Psi^\dagger_j \Psi_j \right) \Psi_i \, , \\
\sigma^z_i &= 2\, \Psi^\dagger_i \Psi_i-1 \, . \nonumber
\end{align}
In terms of the fermionic operators $\Psi^\dagger_i$ and $\Psi_i$, the Hamiltonian
(\ref{XX-hamiltonian}) represents free spinless fermions on a one-dimensional lattice
with $J$ being the hopping amplitude and $h$ corresponding to the chemical potential.
The ground state of such a system is a Fermi sea where all the states below
the Fermi wave vector $k_F=\arccos(h/2J)$ are filled (here we assume that $|h|<2J$: otherwise the 
ground state is a trivial fully polarized state corresponding to $k_F=0$ or $k_F=\pi$). 
The parameter $k_F$ ranges between $0$ and $\pi$ and fully describes the ground state.
The average density of fermions in the ground state is $k_F/\pi$, which 
corresponds to $\langle \sigma_{i}^{z}\rangle=2k_{F}/\pi -1$.

We will be interested in the transverse spin correlation function (\ref{CF-main-scaling}).
For our purposes, it will be more convenient to introduce a more general correlation
function involving a ``counting parameter'' $\kappa$:
\begin{equation}
\Sigma(\kappa,k_F,L) = \left\langle \Psi^\dagger_0 
\exp \left(2\pi i \kappa \sum_{1\le j \le L-1} \Psi^\dagger_j \Psi_j \right) \Psi_L \right\rangle\, ,
\label{cf-kappa-definition}
\end{equation}
where the average is taken over the ground state in an infinite system. Then, 
by the Jordan--Wigner transformation, the transverse spin correlation function is
a particular case of this definition:
\begin{equation}
\langle \sigma^+_i \sigma^-_{i+L} \rangle = \Sigma\left(\kappa{=}\frac{1}{2},k_F,L\right)\, .
\label{kappa-one-half}
\end{equation}
Note that $\Sigma(\kappa,k_F,L)$ is explicitly periodic in $\kappa$ with period one,
since the number of fermions $\sum_{1\le j \le L-1} \Psi^\dagger_j \Psi_j$ is integer.

Using the Wick theorem, the correlation function (\ref{cf-kappa-definition})
can be expressed as \cite{lieb:61,schultz:63}
\begin{multline}
\Sigma(\kappa,k_F,L) = \\
= (1-e^{2\pi i\kappa})^{-1} \det_{1\le i,j \le L}
\left[ \left( 1- e^{2\pi i\kappa} \right) a_{i-j+1} - \delta_{i-j+1} \right]\, ,
\label{CF-determinant}
\end{multline}
where
\begin{equation}
a_{i-j}= 
\left\langle \Psi^\dagger_i \Psi_j \right\rangle =
\begin{cases}
\frac{\sin k_F(i-j)}{\pi(i-j)}\, , & i\ne j\, , \\
k_F/\pi \, , & i=j\, 
\end{cases}
\label{matrix-a}
\end{equation}
and
\begin{equation}
\delta_{i-j}= 
\begin{cases}
0 \, , & i\ne j \, , \\
1 \, , & i=j\, 
\end{cases}
\end{equation}

The determinant in Eq.~(\ref{CF-determinant}) is very similar
to that studied in Ref.~\onlinecite{ivanov:13-2} for the correlation function
\begin{multline}
\chi(\kappa,k_F,L) =
\left\langle \exp \left(2\pi i \kappa \sum_{1\le j \le L} \Psi^\dagger_j \Psi_j \right) \right\rangle \\
= \det_{1\le i,j \le L}
\left[ \left( e^{2\pi i\kappa} - 1 \right) a_{i-j} + \delta_{i-j} \right]\, .
\label{counting-determinant}
\end{multline}
Namely, up to an overall sign, the Toeplitz matrices (\ref{CF-determinant})
and (\ref{counting-determinant}) differ only by a shift by one row
(or, equivalently, by one column). The determinants of such Toeplitz
matrices may therefore be related using the Desnanot-Jacobi identity \cite{muir:06,brualdi:83}
(also known as the Lewis Carroll identity \cite{dodgson:66}, which is a particular case of
the Muir relations \cite{muir:83}):
\begin{multline}
[\chi(\kappa,k_F,L)]^2 - [(1-e^{2\pi i\kappa}) \Sigma(\kappa,k_F,L)]^2 \\
= \chi(\kappa,k_F,L-1)\, \chi(\kappa,k_F,L+1)\, .
\label{Dodgson-relation}
\end{multline}
This relation determines $\Sigma(\kappa,k_F,L)$ up to a sign,
once $\chi(\kappa,k_F,L)$ is known. The sign of $\Sigma(\kappa,k_F,L)$
may, in turn, be fixed independently from the known main
asymptotics (\ref{CF-main-scaling}) \cite{note-xL}.

We can now use the asymptotic expansion for $\chi(\kappa,k_F,L)$ 
derived in Ref.~\onlinecite{ivanov:13-2}:
\begin{equation}
\chi(\kappa,k_F,L)=\sum_{j=-\infty}^{+\infty} \chi_* (\kappa+j, k_F, L)\, ,
\label{expansion-1}
\end{equation}
where
\begin{multline}
\chi_* (\kappa, k_F, L) = \exp \Big[
2 i \kappa k_F L - 2 \kappa^2 \ln (2L\sin k_F) \\
 + {\tilde C}(\kappa)
+\sum_{n=1}^{\infty} {\tilde F}_n(\kappa,k_F)\, (i L)^{-n} \Big]\, ,
\label{expansion-2}
\end{multline}
\begin{equation}
{\tilde C}(\kappa) = 2 \ln [G(1+\kappa) G(1-\kappa)]\, ,
\end{equation}
$G()$ is the Barnes G function, and ${\tilde F}_n(\kappa,k_F)$ are
some polynomials in $\kappa$ and $\cot k_F$.

By re-expressing $\Sigma(\kappa,k_F,L)$ from the relation (\ref{Dodgson-relation}),
we arrive at a similar asymptotic expression,
\begin{multline}
\Sigma(\kappa,k_F,L)=\left( 1-e^{2\pi i \kappa} \right)^{-1}
\sqrt{\frac{2\sin k_F}{L}} \\
\times
\sum_{j=-\infty}^{+\infty} \Sigma_* (\kappa+j+\frac{1}{2}, k_F, L)\, ,
\label{sigma-expansion-1}
\end{multline}
where
\begin{multline}
\Sigma_* (\bar\kappa, k_F, L) = \exp \Big[
2 i \bar\kappa k_F L - 2 \bar\kappa^2 \ln (2L\sin k_F) \\
 + C_\Sigma(\bar\kappa)
+\sum_{n=1}^{\infty} H_n(\bar\kappa,k_F)\, (i L)^{-n} \Big]\, ,
\label{sigma-expansion-2}
\end{multline}
and
\begin{equation}
C_\Sigma(\bar\kappa) = 
\ln [G(\frac{3}{2}+\bar\kappa) G(\frac{1}{2}+\bar\kappa) G(\frac{1}{2}-\bar\kappa) G(\frac{3}{2}-\bar\kappa)]\, .
\label{sigma-c}
\end{equation}
We use the notation $\bar\kappa$ for the variables in Eqs.~(\ref{sigma-expansion-2}) and (\ref{sigma-c})
to emphasize that this variable is shifted by a half-integer from the original variable $\kappa$.
The coefficients $H_n(\bar\kappa,k_F)$ can be calculated from the coefficients ${\tilde F}_n(\kappa,k_F)$
order by order. 
We have calculated the first ten orders
using Mathematica software \cite{Wolfram}. The first six coefficients are:
\begin{align}
\label{H-coefficients}
H_1(\bar\kappa,k_F) & = 
   \left(\bar\kappa^2 -\frac{1}{4}\right) 
   \cdot  2\bar\kappa \cot k_F
   \nonumber\\
H_2(\bar\kappa,k_F) & =
   \left(\bar\kappa^2 -\frac{1}{4}\right) 
   \left[ \left(\frac{5}{2}\bar\kappa^2-\frac{1}{8}\right) \cot^2 k_F
   +\frac{4}{3}\bar\kappa^2 \right]
   \nonumber\\
H_3(\bar\kappa,k_F) & = 
   \left(\bar\kappa^2 -\frac{1}{4}\right) 
   \bigg[\left(\frac{11}{2} \bar\kappa^3-\frac{3}{8}\bar\kappa\right) \cot^3 k_F 
   \nonumber\\
 & +\left(\frac{9}{2}\bar\kappa^3-\frac{1}{8}\bar\kappa \right) \cot k_F \bigg]
   \nonumber\\
H_4(\bar\kappa,k_F) & = 
   \left(\bar\kappa^2 -\frac{1}{4}\right)
   \bigg[\left(\frac{63}{4} \bar\kappa^4 -\frac{11}{16} \bar\kappa^2-\frac{1}{16}\right) \cot^4 k_F
   \nonumber \\
 & +\left(\frac{35}{2} \bar\kappa^4 +\frac{1}{8} \bar\kappa^2-\frac{1}{8}\right) \cot^2 k_F
   \nonumber \\
 & +\frac{167}{60} \bar\kappa^4+\frac{29}{80} \bar\kappa^2 -\frac{1}{32} \bigg] 
   \nonumber\\
H_5(\bar\kappa,k_F) & = 
   \left(\bar\kappa^2 -\frac{1}{4}\right) 
   \nonumber \\
 & \times
   \bigg[ \left(\frac{527}{10} \bar\kappa^5+\frac{17}{10} \bar\kappa^3
      -\frac{23}{32} \bar\kappa \right) \cot^5 k_F
   \nonumber \\
 & +\left(74 \bar\kappa^5+7 \bar\kappa^3-\frac{11}{8}\bar\kappa \right) \cot^3 k_F
   \nonumber \\ 
 & +\left(\frac{45}{2} \bar\kappa^5 +\frac{9}{2} \bar\kappa^3
      -\frac{17}{32} \bar\kappa \right) \cot k_F \bigg]
   \nonumber\\
H_6(\bar\kappa,k_F) & = 
   \left(\bar\kappa^2 -\frac{1}{4}\right) 
   \nonumber \\
 & \hspace{-1cm}
   \times
   \bigg[ \left(\frac{3129}{16} \bar\kappa^6 +\frac{6271}{192} \bar\kappa^4
      -\frac{3599}{768} \bar\kappa^2-\frac{539}{3072}\right) \cot^6 k_F
   \nonumber \\
 & \hspace{-1cm}
   +\left(\frac{2655}{8} \bar\kappa^6 +\frac{2563}{32} \bar\kappa^4
       -\frac{1155}{128} \bar\kappa^2 -\frac{223}{512} \right) \cot^4 k_F
   \nonumber \\
 & \hspace{-1cm}
   +\left(\frac{2385}{16} \bar\kappa^6 +\frac{3341}{64} \bar\kappa^4
       -\frac{1021}{256} \bar\kappa^2 -\frac{353}{1024}\right) \cot^2 k_F
   \nonumber \\    
 & \hspace{-1cm}
   +\frac{236}{21} \bar\kappa^6 +\frac{781}{126} \bar\kappa^4
       -\frac{17}{504} \bar\kappa^2 -\frac{1}{16} \bigg] 
\end{align}
The coefficients $H_7(\bar\kappa,k_F)$ to $H_{10}(\bar\kappa,k_F)$ are
listed in the Appendix.

The expansion (\ref{sigma-expansion-1})--(\ref{sigma-expansion-2}), together with
the algorithm for computing the coefficients $H_n(\bar\kappa,k_F)$ [starting from
the coefficients ${\tilde F}_n(\kappa,k_F)$], constitutes the main result of the present work.
The algorithm for calculating ${\tilde F}_n(\kappa,k_F)$ using discrete
Painlev\'e equations was reported earlier in Ref.~\onlinecite{ivanov:13-2}.
Combined together, these results provide an algorithm for calculating the expansion 
for $\Sigma(\kappa,k_F,L)$ to any given order. The transverse spin
correlations can be obtained by setting $\kappa=1/2$ in all the formulas
(so that the summation in the expansion (\ref{sigma-expansion-1})--(\ref{sigma-expansion-2})
is performed over all integer $\bar\kappa$).

We note several properties of the coefficients $H_n(\bar\kappa,k_F)$. 
Similarly to ${\tilde F}_n(\kappa,k_F)$, they are polynomials in 
$\bar\kappa$ and $\cot k_F$ with rational coefficients. These polynomials have
degrees $(n+2)$ and $n$ in $\bar\kappa$ and $\cot k_F$, respectively and are
of a fixed parity in each of these variables (even for even $n$ and odd for odd $n$). 
Moreover, they are all divisible by $(\bar\kappa^2 -1/4)$: this property must persist to all orders, 
since it guarantees that the expansion (\ref{sigma-expansion-1})--(\ref{sigma-expansion-2})
reproduces the fermionic correlation function (\ref{matrix-a}) in the limit $\kappa\to 0$.

\section{Transverse spin correlations in the XX chain}
\label{sec:spin}

We now specify to the case of the transverse spin correlation function 
(\ref{kappa-one-half}). In this case, all the powers of $L^{-1}$ in
the expansion (\ref{sigma-expansion-1})--(\ref{sigma-expansion-2}) are
integer, and the expansion may be rewritten in the form
\begin{multline}
\label{spin-expansion}
\langle \sigma^+_i \sigma^-_{i+L} \rangle =
e^{C_\Sigma(0)} \sqrt{\frac{\sin k_F}{2L}} \\
\times \sum_{j=-\infty}^{+\infty} (2\sin k_F)^{-2j^2}
e^{2i\, j\, k_F L} \sum_{n=0}^{\infty} \frac{\alpha_{jn}(k_F)}{(iL)^n}\, .
\end{multline}
This form of expansion has already been established in Ref.~\onlinecite{vaidya:79}.
The coefficients $\alpha_{jn}(k_F)$ can be calculated in a simple manner from
the coefficients $H_n(\bar\kappa,k_F)$. Note that at any given order $n$,
the coefficients $\alpha_{jn}(k_F)$ are nonzero only for $|j|\le \sqrt{n/2}$.
The explicit form of the coefficients $\alpha_{jn}(k_F)$ for $n$ up to 11 is
given in Appendix.

The beginning of the expansion (\ref{spin-expansion}) reads
\begin{multline}
\label{spin-expansion-third-order}
\langle \sigma^+_i \sigma^-_{i+L} \rangle =
e^{C_\Sigma(0)} \sqrt{\frac{\sin k_F}{2L}} \\
\times \Bigg( 1 - \frac{\cos^2 k_F + 4 \cos (2 k_F L)}{32 (L \sin k_F)^2} \\
-\frac{3 \cos k_F \, \sin(2 k_F L)}{16 (L \sin k_F)^3} + \ldots \Bigg)\, .
\end{multline} 
The leading order gives the asymptotic behavior (\ref{CF-main-scaling}) 
with the correct coefficient \cite{vaidya:79,jimbo:80,creamer:81,gangardt:04:06}:
\begin{equation}
	e^{C_\Sigma(0)} =  \left[G(1/2)G(3/2)\right]^{2} = 2^{1/6}e^{1/2}A^{-6}\, ,
\end{equation}
where $A = 1.2824271291\ldots$ is the Glaisher-Kinkelin constant.
The subsequent coefficients $\alpha_{jn}$ reproduce, in particular, the corrections calculated
in Refs.~\onlinecite{creamer:81,gangardt:04:06}. 

\section{Numerical illustration}
\label{sec:numerics}

\begin{figure}
\centerline{\includegraphics[width=.48\textwidth]{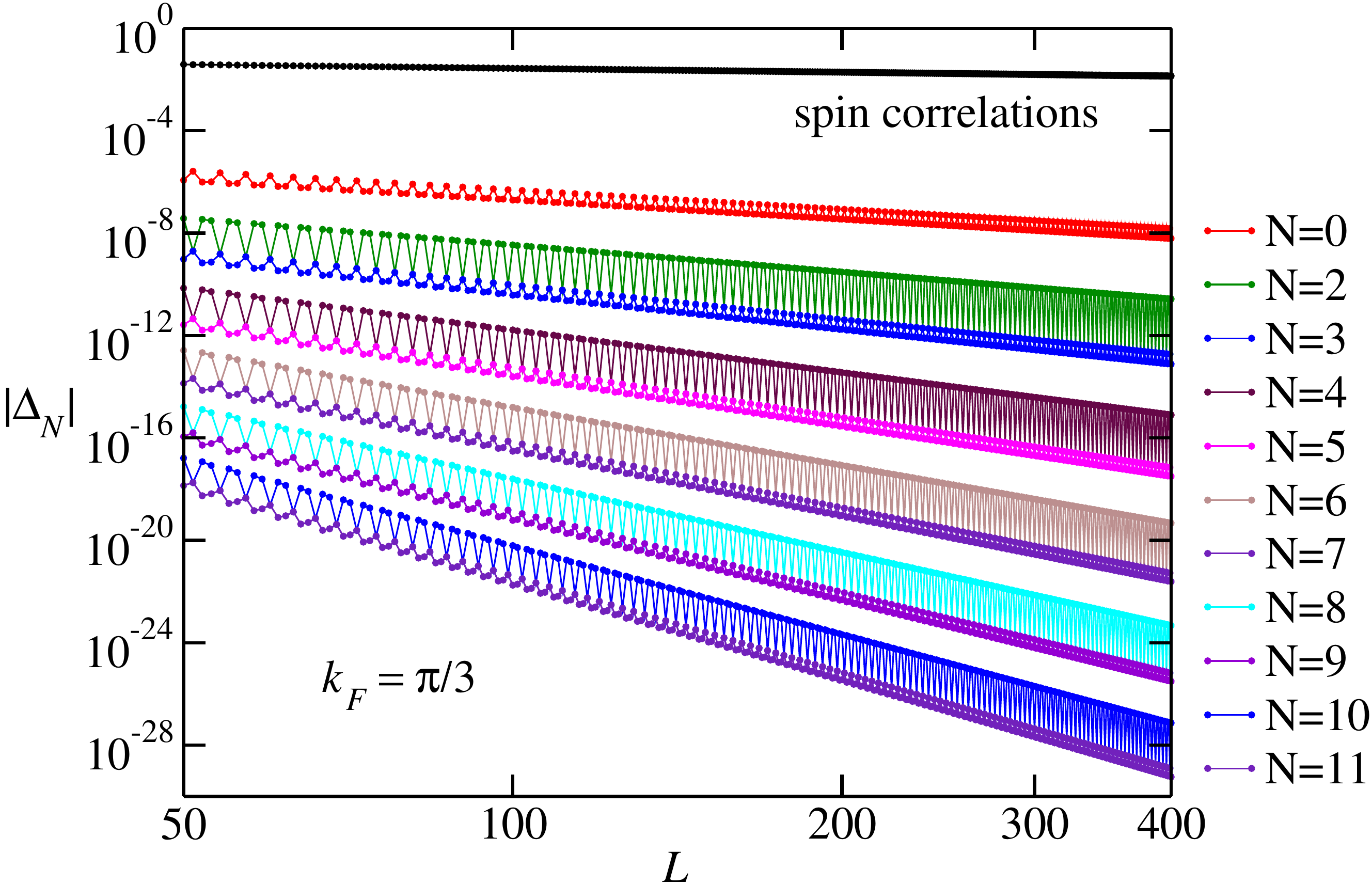}}
\caption{The difference $\Delta_N$ between the left-hand side of Eq.~(\ref{spin-expansion})
and its right-hand side with the sum over $n$ restricted to
$n\le N$ is plotted as a function of $L$. The value of $k_F$ is $\pi/3$. The upper 
line is the spin correlations [the left-hand side of Eq.~(\ref{spin-expansion})], 
and the other data correspond,
top to bottom, to $N=0$ to $N=11$ [excluding $N=1$, since there are no 
first-order terms in the expansion (\ref{spin-expansion})].}
\label{fig:Delta}
\end{figure}

We illustrate our analytic calculation with a numerical example of the correlation 
function (\ref{kappa-one-half}). We have chosen the Fermi wave vector $k_F=\pi/3$
(corresponding to the $z$ polarization equal to $1/3$ of the full polarization) and
have numerically calculated the corresponding determinants for distance
$L$ up to 400. In our numerics, we have used the LAPACK library \cite{lapack:99} compiled 
to work with 128-bit floating-point numbers, together with the quadmath C library.

In Fig.~\ref{fig:Delta} we plot the difference $\Delta_N (k_F,L)$ 
between the left-hand side
of Eq.~(\ref{spin-expansion}) and its right-hand side with the sum over $n$
restricted to $n \le N$. These results show that, even though our analytical
calculations involved a non-rigorous analytic continuation of the asymptotic series
to half-integer values of $\kappa$, such an analyticity, in fact, holds.
A similar conclusion was also reached in 
Refs.\ \onlinecite{calabrese:10,abanov:11,suesstrunk:12,ivanov:13-2}
for the expansion of $\chi(\kappa, k_F,L)$.

\section{Discussion}
\label{sec:discussion}

In the present paper, we apply the earlier results of Ref.~\onlinecite{ivanov:13-2}
on the Toeplitz determinants with the sine kernel to deriving a Fisher--Hartwig
expansion for the correlation function (\ref{cf-kappa-definition}) [including,
the transverse spin correlations (\ref{kappa-one-half}) as a particular case].
The expansion is not rigorously proven and remains a conjecture supported by
several arguments.

Away from the line $\re \kappa = j+1/2$ (with an integer $j$), this expansion may
be verified order by order using the methods of Refs.~\onlinecite{calabrese:10,ivanov:13-2}
(and of Ref.~\onlinecite{ivanov:13-1} in the continuous limit $k_F \to 0$).
The verification was actually performed to the tenth order in the lattice case
and to the fifteenth order in the continuous limit, and this leaves little doubt
about the validity of the general form of the expansion to all orders.

On the line $\re \kappa = j+1/2$ (relevant for the case of the transverse
correlations in the XX model), the situation is more delicate. In this case,
the expansion cannot even be rigorously derived to any order, but is obtained
by an analytic continuation from other values of $\kappa$. This is not
a mathematically justified procedure, and therefore our results at $\re \kappa = j+1/2$
are additionally based on the assumption that the expansion
(\ref{expansion-1})--(\ref{expansion-2}) of $\chi(\kappa,k_F,L)$ is 
analytically continuable, term by term, across the line 
$\re \kappa = j+1/2$ (see Refs.\ \onlinecite{ivanov:13-1,ivanov:13-2}). 
The corresponding analytic continuation for the expansion 
(\ref{sigma-expansion-1})--(\ref{sigma-expansion-2}) of $\Sigma(\kappa,k_F,L)$
follows from this assumption, together with the Lewis Carroll identity
(\ref{Dodgson-relation}). At this point it is not clear how to prove 
this assumption. However, available numerical
studies (Refs.\ \onlinecite{calabrese:10,abanov:11,suesstrunk:12,ivanov:13-2} 
and this paper) indicate that, in the examples and to the orders considered,
the analytic continuation of the expansions to the line $\re \kappa = j+1/2$
is indeed possible.

These conjectures present a challenge to future mathematical studies of
Toeplitz determinants. Besides proving them, an interesting question remains
if they are valid for other Toeplitz determinants with
Fisher--Hartwig singularities, or, even more generally, for pseudo-differential
operators with discontinuous symbols \cite{sobolev:10}. Transferring some
of the results on the Fisher--Hartwig expansion to spectral properties
of such operators would have implications in extending the
Widom conjecture \cite{widom:82} to a wider class of functions.
In particular, this may lead to extracting subleading corrections to the
von Neumann entanglement entropy for free fermions in higher dimensions (similarly
to the one-dimensional case \cite{suesstrunk:12,ivanov:13-2}).

Another use of the present results may be in application to one-dimensional
bosonization (describing the low-energy fermionic degrees of freedom in terms
of bosonic fields) \cite{stone:94}. While the subleading bosonization 
terms (responsible for the discreteness of fermionic particles) are model dependent\cite{haldane:81},
it might be possible to fix them for the specific model (free fermions on a chain)
by using expansions for correlation functions obtained from Toeplitz
determinants. 

\begin{acknowledgments}
We thank V.~Fock for bringing to our attention the Lewis Carroll identity for determinants.
The work of AGA was supported by the NSF under grant no.\ DMR-1206790.
\end{acknowledgments}

\section*{Appendix}
\label{sec:Appendix}

The coefficients $H_n(\bar\kappa,k_F)$ in orders seven to ten are:
\begin{align}
H_7(\bar\kappa,k_F) & =    
\left(\bar\kappa^2 -\frac{1}{4}\right) 
   \nonumber \\
 & \hspace{-1.5cm}
   \times
   \bigg[ \bigg(\frac{175045}{224} \bar\kappa^7 + \frac{257105}{896} \bar\kappa^5 
   \nonumber\\
 & -\frac{51889}{3584} \bar\kappa^3 -\frac{8243}{2048} \bar\kappa\bigg) \cot ^7 k_F
   \nonumber\\
 & \hspace{-1.5cm}  
   +\bigg(\frac{49755}{32} \bar\kappa^7 +\frac{90319}{128} \bar\kappa^5
   \nonumber\\
 & -\frac{10991}{512} \bar\kappa^3 - \frac{21563}{2048} \bar\kappa \bigg) \cot^5 k_F
   \nonumber\\
 & \hspace{-1.5cm}  
   +\bigg(\frac{88735}{96} \bar\kappa^7 +\frac{202627}{384} \bar\kappa^5 
   \nonumber\\
 & -\frac{35}{1536} \bar\kappa^3 -\frac{18261}{2048} \bar\kappa \bigg) \cot^3 k_F
   \nonumber\\
 & \hspace{-1.5cm}  
   +\bigg(\frac{4765}{32} \bar\kappa^7 +\frac{14281}{128} \bar\kappa^5 
   +\frac{3031}{512} \bar\kappa^3 -\frac{4669}{2048} \bar\kappa \bigg) \cot k_F \bigg]\, ,
   \nonumber\\
H_8(\bar\kappa,k_F) & =    
\left(\bar\kappa^2 -\frac{1}{4}\right) 
   \nonumber \\
 & \hspace{-1.5cm}
   \times   
   \bigg[ \bigg(\frac{422565}{128} \bar\kappa^8 +\frac{1079589}{512} \bar\kappa^6
   +\frac{235859}{2048} \bar\kappa^4
   \nonumber\\
 & -\frac{407593}{8192} \bar\kappa^2
   -\frac{10045}{8192} \bigg) \cot^8 k_F
   \nonumber\\
 & \hspace{-1.5cm}
   +\bigg(\frac{723283}{96} \bar\kappa^8 +\frac{2134807}{384} \bar\kappa^6
   +\frac{659881}{1536} \bar\kappa^4 
   \nonumber\\
 & -\frac{281377}{2048} \bar\kappa^2
   -\frac{1945}{512}\bigg) \cot^6 k_F 
   \nonumber\\
 & \hspace{-1.5cm}
   +\bigg(\frac{356807}{64} \bar\kappa^8 +\frac{1234703}{256} \bar\kappa^6
   +\frac{545033}{1024} \bar\kappa^4 
   \nonumber\\
 & -\frac{520155}{4096} \bar\kappa^2
   -\frac{17025}{4096}\bigg) \cot^4 k_F 
   \nonumber\\
 & \hspace{-1.5cm}
   +\bigg(\frac{44825}{32} \bar\kappa^8 +\frac{186653}{128} \bar\kappa^6
   +\frac{121011}{512} \bar\kappa^4 
   \nonumber\\
 & -\frac{83617}{2048} \bar\kappa^2 
   -\frac{1865}{1024}\bigg) \cot^2 k_F 
   \nonumber\\
 & \hspace{-1.5cm}
   +\frac{353777}{5760} \kappa^8 +\frac{1874177}{23040} \bar\kappa^6
   +\frac{1954679}{92160} \bar\kappa^4
   \nonumber\\
 & -\frac{252007}{122880} \bar\kappa^2
   -\frac{1697}{8192} \bigg]\, , 
   \nonumber\\
H_9(\bar\kappa,k_F) & =    
\left(\bar\kappa^2 -\frac{1}{4}\right) 
   \nonumber\\
 & \hspace{-1.5cm}
   \times      
   \bigg[\bigg(\frac{1398251}{96} \bar\kappa^9 +\frac{2759869}{192} \bar\kappa^7
   +\frac{263645}{96} \bar\kappa^5
   \nonumber\\
 & -\frac{383007}{1024} \bar\kappa^3 -\frac{365609}{8192} \bar\kappa \bigg) \cot^9 k_F
   \nonumber\\
 & \hspace{-1.5cm}
   +\bigg(\frac{149997}{4} \bar\kappa^9 +\frac{329653}{8} \bar\kappa^7
   +\frac{144741}{16} \bar\kappa^5
   \nonumber\\
 & -\frac{135021}{128} \bar\kappa^3 -\frac{150545}{1024} \bar\kappa \bigg) \cot^7 k_F
   \nonumber\\
 & \hspace{-1.5cm}
   +\bigg(\frac{2656689}{80} \bar\kappa^9 +\frac{6574367}{160} \bar\kappa^7
   +\frac{424443}{40} \bar\kappa^5
   \nonumber\\
 & -\frac{2573727}{2560} \bar\kappa^3 -\frac{719261}{4096} \bar\kappa \bigg) \cot^5 k_F
   \nonumber\\
 & \hspace{-1.5cm}
   +\bigg( 11414 \bar\kappa^9+\frac{64723}{4} \bar\kappa^7
   +\frac{81147}{16} \bar\kappa^5 
   \nonumber \\
 & -\frac{21975}{64} \bar\kappa^3-\frac{22669}{256} \bar\kappa \bigg) \cot^3 k_F
   \nonumber\\
 & \hspace{-1.5cm}
   +\bigg(\frac{36597}{32} \bar\kappa^9 +\frac{122547}{64} \bar\kappa^7
   +\frac{24549}{32} \bar\kappa^5 
   \nonumber\\
 & -\frac{21459}{1024} \bar\kappa^3-\frac{123653}{8192} \bar\kappa \bigg) \cot k_F \bigg]\, ,
   \nonumber\\   
H_{10}(\bar\kappa,k_F) & =    
\left(\bar\kappa^2 -\frac{1}{4}\right) 
   \nonumber\\
 & \hspace{-1.5cm}
   \times  \bigg[\bigg(\frac{266149}{4} \bar\kappa^{10} +\frac{60326939}{640} \bar\kappa^8
   +\frac{44195357}{1280} \bar\kappa^6
   \nonumber\\
 & \hspace{-0.5cm}
   -\frac{181763}{320} \bar\kappa^4 -\frac{17390113}{20480} \bar\kappa^2
   -\frac{2645371}{163840} \bigg) \cot^{10} k_F
   \nonumber\\
 & \hspace{-1.5cm}
   +\bigg(\frac{381511}{2} \bar\kappa^{10} +\frac{588149}{2} \bar\kappa^8
   +\frac{15122545}{128} \bar\kappa^6 
   \nonumber\\
 & \hspace{-0.5cm}
   -\frac{142507}{512} \bar\kappa^4-\frac{6068297}{2048} \bar\kappa^2
   -\frac{494569}{8192}\bigg) \cot^8 k_F 
   \nonumber\\
 & \hspace{-1.5cm}
   +\bigg(\frac{786877}{4} \bar\kappa^{10} +\frac{21302633}{64} \bar\kappa^8
   +\frac{19058315}{128} \bar\kappa^6
   \nonumber\\
 & +\frac{9023}{4} \bar\kappa^4-\frac{7867861}{2048} \bar\kappa^2
   -\frac{1402057}{16384} \bigg) \cot^6 k_F
   \nonumber\\
 & \hspace{-1.5cm}
   +\bigg(\frac{344253}{4} \bar\kappa^{10} +\frac{5177563}{32} \bar\kappa^8
   +\frac{657703}{8} \bar\kappa^6 
   \nonumber\\
 & +\frac{810981}{256} \bar\kappa^4-\frac{2264409}{1024} \bar\kappa^2
   -\frac{457181}{8192}\bigg) \cot^4 k_F
   \nonumber\\
 & \hspace{-1.5cm}
   +\bigg(\frac{27937}{2} \bar\kappa^{10} +\frac{3807047}{128} \bar\kappa^8
   +\frac{4528105}{256} \bar\kappa^6
   \nonumber\\
 & +\frac{5181}{4} \bar\kappa^4 -\frac{2064369}{4096} \bar\kappa^2 
   -\frac{512839}{32768} \bigg) \cot^2 k_F
   \nonumber\\
 & \hspace{-1.5cm}  
   +\frac{264031}{660} \bar\kappa^{10} +\frac{1055183}{1056} \bar\kappa^8
   +\frac{15531919}{21120} \bar\kappa^6
   \nonumber\\
 & +\frac{2831703}{28160} \bar\kappa^4 -\frac{492119}{22528} \bar\kappa^2
   -\frac{2301}{2048} \bigg]\, .
\end{align}

We also list all nonzero coefficients $\alpha_{jn}(k_F)$ for $n\le 11$.
Only coefficients with $j\ge 0$ are presented because of the symmetry 
$\alpha_{-jn}(k_F) = (-1)^n \alpha_{jn} (k_F)$:
\begin{align}
\alpha_{00} (k_F) &= 1 \, ,
  \qquad
\alpha_{02} (k_F) = \frac{1}{32} \cot^2 k_F \, , 
  \nonumber\\
\alpha_{12} (k_F) &= \frac{1}{4} \, ,
  \qquad
\alpha_{13} (k_F) = \frac{3}{8} \cot k_F\, ,
  \nonumber\\
\alpha_{04} (k_F) &= \frac{33}{2048} \cot^4 k_F + \frac{1}{32} \cot^2 k_F + \frac{1}{128}\, ,
  \nonumber\\
\alpha_{14} (k_F) &= \frac{93}{128} \cot^2 k_F + \frac{1}{4}\, ,
  \nonumber\\
\alpha_{15} (k_F) &= \frac{453}{256} \cot^3 k_F + \frac{153}{128} \cot k_F\, ,
  \nonumber\\
\alpha_{06} (k_F) &= \frac{2907}{65536} \cot^6 k_F + \frac{225}{2048} \cot^4 k_F 
  \nonumber\\
 &+ \frac{177}{2048} \cot^2 k_F + \frac{1}{64} \, ,
  \nonumber\\
\alpha_{16} (k_F) &= \frac{42633}{8192} \cot^4 k_F + \frac{1341}{256} \cot^2 k_F 
+ \frac{363}{512}\, ,
  \nonumber\\
\alpha_{17} (k_F) &= \frac{293895}{16384} \cot^5 k_F +\frac{98325}{4096} \cot^3 k_F
  \nonumber\\
  & \hspace{1cm}
+\frac{7011}{1024} \cot k_F\, ,
  \nonumber\\
\alpha_{08} (k_F) &= \frac{2584107}{8388608} \cot^8 k_F +\frac{1953}{2048} \cot^6 k_F
  \nonumber\\
  & \hspace{-0.5cm}
+\frac{273267}{262144} \cot^4 k_F +\frac{467}{1024} \cot^2 k_F +\frac{849}{16384} \, , 
  \nonumber\\
\alpha_{18} (k_F) &= \frac{18582471}{262144} \cot^6 k_F +\frac{970029}{8192} \cot^4 k_F
  \nonumber\\
  & \hspace{0.5cm}
+\frac{428337}{8192} \cot^2 k_F + \frac{1985}{512}\, ,
  \nonumber\\
\alpha_{28} (k_F) &= \frac{9}{256}\, ,
  \nonumber\\
\alpha_{19} (k_F) &= \frac{165603555}{524288} \cot^7 k_F +\frac{165851145}{262144} \cot^5 k_F
  \nonumber\\
  & \hspace{0.5cm}
+\frac{12465711}{32768} \cot^3 k_F +\frac{2052147}{32768} \cot^k_F\, ,
  \nonumber\\
\alpha_{29} (k_F) &= \frac{135}{256} \cot k_F\, ,
  \nonumber\\
\alpha_{0,10} (k_F) &= 
  \frac{1086306255}{268435456} \cot^{10} k_F +\frac{63442485}{4194304} \cot^8 k_F
  \nonumber\\
  & \hspace{0.5cm}
  +\frac{44944731}{2097152} \cot^6 k_F +\frac{3662175}{262144} \cot^4 k_F
  \nonumber\\
  & \hspace{0.5cm}
  +\frac{1026407}{262144} \cot^2 k_F +\frac{1151}{4096}\, ,
  \nonumber\\
\alpha_{1,10} (k_F) &= \frac{52562685915}{33554432} \cot^8 k_F
  \nonumber\\
  & \hspace{-0.5cm}
+\frac{1918340595}{524288} \cot^6 k_F + \frac{2927324961}{1048576} \cot^4 k_F
  \nonumber\\
  & \hspace{-0.5cm}
+ \frac{48020019}{65536} \cot^2 k_F + \frac{4529519}{131072} \, ,
  \nonumber\\
\alpha_{2,10} (k_F) &= \frac{43065}{8192} \cot^2 k_F + \frac{45}{64}\, ,
  \nonumber\\
\alpha_{1,11} (k_F) &= \frac{574231373145}{67108864} \cot^9 k_F 
  \nonumber\\
  & \hspace{-0.5cm}  
+ \frac{191554621335}{8388608} \cot^7 k_F + \frac{44340529815}{2097152} \cot^5 k_F
  \nonumber\\
  & \hspace{-0.5cm} 
   + \frac{8076485685}{1048576} \cot^3 k_F + \frac{219859155}{262144} \cot k_F \, ,
  \nonumber\\
\alpha_{2,11} (k_F) &= \frac{368685}{8192} \cot^3 k_F + \frac{62505}{4096} \cot k_F\, .
\end{align}


\begin{thebibliography}{99}
\bibitem{BosLead}
Probably, the easiest way of deriving the leading term (\ref{CF-main-scaling}) is to use the bosonization 
technique \cite{luther:75}.
\bibitem{note-vaidya}
The expansion coefficients in Refs.~\onlinecite{vaidya:79,jimbo:80} contain typos
which were corrected later in Refs.~\onlinecite{creamer:81,gangardt:04:06}.
\bibitem{note-xL}
Remarkably, the quantity $x_L$ studied in Refs.\ \onlinecite{calabrese:10,ivanov:13-2}
and obeying the discrete version of Painlev\'e equations is given by the ratio of the
two determinants: $x_L = (1-e^{2\pi i \kappa})\Sigma(\kappa,k_F,L)/\chi(\kappa,k_F,L)$.
\end{thebibliography}
\end{document}